\documentclass[preprint,3p,times,twocolumn]{elsarticle}

\usepackage{amssymb}
\usepackage{amsmath}
\usepackage{hhline}
\usepackage{colortbl}
\usepackage{graphicx}
\usepackage{bm}
\usepackage{stfloats}
\usepackage[ruled, vlined, linesnumbered]{algorithm2e}
\usepackage{multirow}

\journal{Expert Systems With Applications}

\begin{document}

\begin{frontmatter}



\title{A Novel Deep Reinforcement Learning Based Automated Stock Trading System Using Cascaded LSTM Networks}


\author[1]{Jie Zou}
\ead{ianzou2000@163.com}

\author[1]{Jiashu Lou}
\ead{loujiashu@163.com}

\author[1]{Baohua Wang\corref{cor1}}
\ead{bhwang@szu.edu.cn}

\author[1]{Sixue Liu}
\ead{18801117020@163.com}

\cortext[cor1]{Corresponding author}

\affiliation[1]{organization={College of Mathematics and Statistics},
            addressline={Shenzhen University}, 
            postcode={518060}, 
            state={Guangdong},
            country={China}}

\begin{abstract}
More and more stock trading strategies are constructed using deep reinforcement learning (DRL) algorithms, but DRL methods originally widely used in the gaming community are not directly adaptable to financial data with low signal-to-noise ratios and unevenness, and thus suffer from performance shortcomings. In this paper, to capture the hidden information, we propose a DRL based stock trading system using cascaded LSTM (CLSTM-PPO Model), which first uses LSTM to extract the time-series features from daily stock data, and then the features extracted are fed to the agent for training, while the strategy functions in reinforcement learning also use another LSTM for training. Experiments in 30 stocks from the Dow Jones Industrial index (DJI) in the US, 30 stocks from SSE50 on the Shanghai Stock Exchange in China, 30 stocks from SENSEX on the Bombay Stock Exchange in India and 30 stocks from FTSE 100 on London Stock Exchange in the UK show that our model outperforms previous baseline models in terms of cumulative returns by 5\% to 52\%, maximum earning rate by 8\% to 52\%, average profitability per trade by 6\% to 14\% and these advantages are more significant in the Chinese stock market, an emerging market, where cumulative returns have improved by 84.4\% and the Sharpe ratio by 37.4\% than ensemble strategy. It indicates that our proposed method is a promising way to build a automated stock trading system.
\end{abstract}




\end{frontmatter}


\section{Introduction}
In recent years, more and more institutional and individual investors are using machine learning and deep learning methods for stock trading and asset management, such as stock price prediction using Random Forests, Long Short-Term Memory(LSTM) Neural Networks or Support Vector Machines\cite{1}, which help traders to get well-performing online strategies and obtain higher returns than strategies using only traditional factors\cite{2}\cite{3}\cite{4} .

However, there are three main limitations of machine learning methods for stock market prediction: (i) financial market data are filled with noise and are unstable, and also contain the interaction of many unmeasurable factors. Therefore, it is very difficult to take into account all relevant factors in complex and dynamic stock markets\cite{5}\cite{6}\cite{7}. (ii) Stock prices can be influenced by many other factors, such as political events, the behavior of other stock markets or even the psychology of investors\cite{8}. (iii) Most of the methods are performed based on supervised learning and require training sets that are labeled with the state of the market, but such machine learning classifiers are prone to suffer from overfitting, which reduces the generalization ability of the model\cite{9}.

Fundamental data from financial statements and other data from business news, etc. are combined with machine learning algorithms that can obtain investment signals or make predictions about the prospects of a company\cite{10}\cite{11}\cite{12}\cite{13} to screen for good investment targets. Such an algorithm solves the problem of stock screening, but it cannot solve how to allocate positions among the investment targets. In other words, it is still up to the trader to judge the timing of entry and exit.

Besides using some machine learning and deep learning methods to predict stock prices, there are some classical models like Geometric Brownian Motion. Agustini\cite{50} demonstrated that Geometric Brownian Motion model has a high accuracy in predicting stock prices and is proven with forecast Mean Absolute Percentage Error(MAPE) value $\le$ 20\% in the Indonesian market. Also, it is a sufficiently simple and highly interpretable model that can be used to predict the stock price as long as we have the historical prices of the stock. However, it also has some limitations: 1. Returns of stocks should follow a log-normal distribution, which may not always be the case. 2. GBM may not capture the complex patterns and trends that can be observed in the financial markets, so its accuracy may be limited when the time span of historical stock price series is too long or when many stocks need to be traded simultaneously. Deep learning, although sometimes lacking in interpretability, is able to capture nonlinear patterns in financial time series, which may be difficult or impossible to model using traditional statistical methods. to solve the problem of automatically trading a portfolio containing dozens of stocks.

To overcome the main limitations listed above, in this paper we use a deep reinforcement learning approach, a branch of deep learning, to construct low-frequency automated stock trading strategies to solve the problem of automatically trading a portfolio containing dozens of stocks with maximizing expected returns. We consider stock trading as a Markov decision process that will be represented by states, actions, rewards, strategies, and values in a reinforcement learning algorithm. Instead of relying on labels (e.g., up and down in the market) to learn, the reinforcement learning approach learns how to maximize the objective function, which can be achieved by maximizing the value function here, during the training phase. We mainly use the PPO algorithm to train the agent and combine it with LSTM to extract time-series features for an initial state of a certain time window length, while the initial state is represented by the adjusted stock price, available balance, number of shares in the underlying asset and some technical indicators: the Moving Average Convergence Divergence (\textbf{MACD}) that measures the difference between two moving averages and uses this information to identify potential trends and changes in momentum in a security or asset, the Relative Strength Index (\textbf{RSI}) that measures the strength of a security or asset's price action by comparing its average gains to its average losses over a specified period of time, the Commodity Channel Index (\textbf{CCI}) that measures the difference between the current price of a security or asset and its average price over a specified period of time and the Average Directional Index (\textbf{ADX}) that measures the strength of a security or asset's trend, illustrated in Figure 1. 

We take the ensemble strategy proposed by Yang, Liu\cite{14} as the baseline and further expand on their work, so the training environment, state space, behavior space and value function we use are consistent with it.

\begin{figure}[htbp!]
	\centering
	\includegraphics[width=1.0\linewidth]{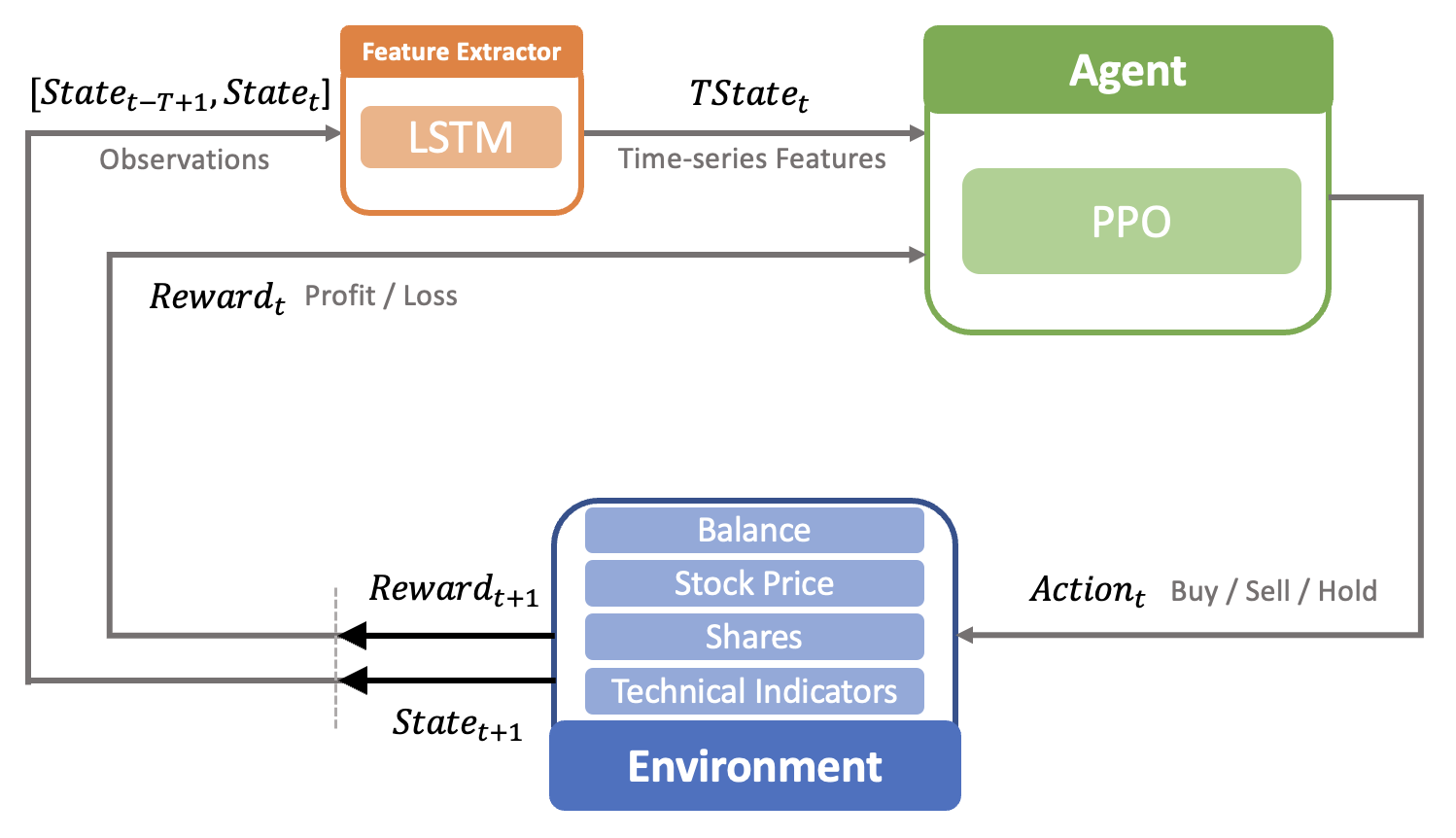}
	\caption{Agent-environment interaction in reinforcement learning}
	\label{ae}
\end{figure}

Experiments show that the automated stock trading strategy based on LSTM outperforms the ensemble strategy and the buy-and-hold strategy represented by the Dow Jones Industrial index (DJI) in terms of cumulative return, and also has better performance in Chinese stock market in terms of cumulative return and Sharpe ratio.

The main contributions of this paper are two-fold: (i) In the literature, security’s past price and related technical indicators are often used to represent state spaces. Instead of using the raw data, we use LSTM to extract the time-series features from stock daily data to  represent state spaces, since the memory property of the LSTM can discover features of the stock market that change over time and integrate hidden information in the time dimension, thus making it more likely that the partially observable Markov decision process (POMDP) is closer to the Markov decision process (MDP). (ii) Different from the previous DRL based methods which use multi-layer neural networks or convolutional networks in agent training, we use LSTM as the training network because it is a type of recurrent neural network capable of learning order dependence in sequence prediction problems. It should be noted that higher-order Markov models can also capture dependencies between a current state and a longer history of past states, which can potentially improve the accuracy of predictions in some scenarios. However, using higher-order models can also come with several challenges and drawbacks. One major challenge is the higher computational cost of training and using higher-order models than LSTM networks, as the number of possible state sequences grows exponentially with the order of the model. This can make it difficult or even infeasible to train and use such models with large datasets or real-time decision-making systems. Another challenge is the potential for overfitting, as higher-order models can become overly complex and capture noise or spurious patterns in the data. This can lead to poor generalization to new data and reduce the model's effectiveness.

The rest of this paper is organized as follows. Section 2 is a brief introduction to the work related to stock trading using reinforcement learning, categorized by the training algorithm. Section 3 focuses on our algorithm, defining the necessary constraints in the environment and the framework of the agent. Section 4 presents the results and analysis of the experiments, covering the introduction of datasets, baseline models and evaluation metrics, process of finding the well-performing parameters, and experiments in Chinese, Indian, UK and US markets. In Conclusion, we summarize the whole work and give directions of improvement for the future.

\section{Related Work}
This section briefly summarizes the application of reinforcement learning, LSTM and some state-of-the-art models in quantitative trading, reviewing three learning methods in reinforcement learning that are frequently applied to financial markets and LSTM neural networks that are applied to predict stock prices. These three learning methods are: critic-only learning, actor-only learning and actor-critic learning.

\subsection{Critic-only}
The Critic-only approach, the most common of the three, uses only the action-value function to make decisions with the aim of maximizing the expected reward for each action choice given the current state. The action-value function Q receives the current state and the possible actions to be taken as input, then outputs an expected Q value as a reward. One of the most popular and successful approaches is Deep Q Network (DQN)\cite{15} and its extensions\cite{16}. Chen\cite{17}, Dang\cite{18} and Jeong\cite{19} used this method to train agents on a single stock or asset. Chen\cite{17}, Huang\cite{20} used Deep Recurrent Q Network (DRQN) trained agents to achieve higher cumulative performance on quantitative trading than baseline models and DQN to achieve higher cumulative returns than baseline models. However, the main limitation of the method is that it performs well on discrete state spaces, but the stock prices are continuous. If a larger number of stocks or assets are selected, the state space and action space will grow exponentially\cite{21}, which will weaken the performance of DQN.

\subsection{Actor-only}
The actor-only approach is able to learn policies directly and the action space can be considered as continuous. Therefore, the advantage of this approach is that it can directly learn the good mapping from a particular state to an action, which can be either discrete or continuous. Its disadvantages are that it requires a large amount of data for experiments and a long time to obtain the well-performing strategy\cite{22}. Deng\cite{23} used this method and applied Recurrent Deep Neural Network to real-time financial trading for the first time. Wu\cite{24} also explored the actor-only method in quantitative trading, where he compared deep neural networks (LSTM) with fully connected networks in detail and discussed the impact of some combinations of technical indicators on the daily data performance of the Chinese market, proving that deep neural networks are superior. The results of his experiments are mixed, and he shows that the proposed approach can yield decent profits in some stocks, but performs mediocrely in others.

\subsection{Actor-critic}
The actor-critic approach aims to train two models simultaneously, with the actor learning how to make the agent respond in a given state and the critic evaluating the responses. Currently, this approach is considered to be one of the most successful algorithms in RL, while Proximal Policy Optimization (PPO) is the most advanced actor-critic approach available. It performs better because it solves the well-known problems when applying RL to complex environments, such as instability due to the distribution of observations and rewards that constantly change as the agent learns\cite{25}. In this paper, the baseline model\cite{14} is constructed based on the actor-critic approach, using a combination of three DRL algorithms: PPO, A2C, and DDPG. However, the agent that learns only with PPO outperforms the resemble strategy in terms of cumulative return.

\subsection{LSTM in stock system}
Although Long Short-Term Memory Networks (LSTM)\cite{26} are traditionally used in natural language processing, many recent works have applied them in financial markets to filter some noise in raw market data\cite{41}\cite{42}\cite{43}\cite{44}. Stock prices and some technical indicators generated from stock prices are interconnected, so LSTM can be used as a feature extractor to extract potentially profitable patterns in the time series of these indicators. Zhang\cite{22}, Wu\cite{24} have tried to integrate LSTM for feature extraction while training agent using DRL algorithm and the experiments have shown that it works better than baseline model. And the work of Lim\cite{45} shows that LSTM delivers superior performance on modelling daily financial data.

\subsection{Some advanced models in quantitative investment}
In addition to using reinforcement learning(RL) for quantitative investing, there are many other innovative approaches. Wu\cite{46} proposed a system based on fuzzy analysis methods, which can classify stocks suited  to momentum- or contrarian-type strategies and it increased the profitability by 1.5 times on the Taiwan 50 dataset. Syu\cite{47} introduced the TripleS, a stock selection system that utilizes fuzzy-set theory to establish the link between stocks and investment strategies. Besides using LSTM to extract time-series features of stock price series, some scholars also represent the stock and its leading indicators (futures/options) price series as graph data, and then use CNN to extract features. Wu\cite{48} presents a two-dimensional tensor input data and feature extraction method for training a CNN network to predict the stock market, which outperforms previous algorithms in avoiding noise and overfitting. Wu\cite{49} proposed a new framework based on CNN and LSTM to predict the direction of the stock market by aggregating multiple variables, automatically extracting features through CNN, and inputting them into LSTM. HIST\cite{51} is a high-frequency trading simulator developed by Microsoft Research Asia. It is designed to provide a realistic environment for developing and testing high-frequency trading algorithms. Qlib\cite{51} which is an open-source Python library developed by Microsoft Research Asia is built on top of HIST and other market simulators and supports a variety of deep learning, reinforcement learning, and traditional machine learning models.
\section{Our method}
\subsection{Stock Market Environment}
The stock market environment used in this paper is a simulation environment developed in Yang\cite{14} based on the OpenAI gym\cite{28}\cite{29}\cite{30}, which is able to give the agent various information for training, such as current stock prices, shareholdings and technical indicators. We use Markov Decision Process (MDP) to model stock trading\cite{36}, so the information that should be included in this multi-stock trading environment are: state, action, reward, policy and Q-value. Then, 30 stocks which are chosen randomly from a stock index in a financial market will be introduced as objects of transaction. Agent will only foucus on these 30 stocks' information, and take actions like buying, selling and holding.

\subsubsection{State Space}
A 181-dimensional vector consisting of seven parts represents the state space of the multi-stock trading environment for these 30 stocks: $[b_t, \bm{p}_t,\bm{h}_t,\bm{M}_t,\bm{R}_t,\bm{C}_t,\bm{X}_t]$. Each of these components is defined as follows. In Yang\cite{14}, this 181-dimensional vector is fed as a state directly into the reinforcement learning algorithm for learning. However, our approach is to give $T$($T$ is the time window of LSTM) 181-dimensional vectors to the LSTM for learning first, and the feature vector generated by the LSTM are given to the agent for learning as our state.

\begin{enumerate}
	\item $b_t\in\mathbb{R}_+$: available balance at current time step t.
	\item $\bm{p}_t\in\mathbb{R}_+^{30}$: adjusted close price of each stock at current time step t.
	\item $\bm{h}_t\in\mathbb{Z}_+^{30}$: number of shares owned of each stock at current time step t.
	\item $\bm{M}_t\in\mathbb{R}^{30}$: Moving Average Convergence Divergence (MACD) is calculated using close price of each stock at current time step t. MACD is one of the most commonly used momentum indicators that identifies moving averages\cite{32}.
	\item $\bm{R}_t\in\mathbb{R}_+^{30}$: Relative Strength Index (RSI) is calculated using close price of each stock at current time step t. RSI quantifies the extent of recent price changes\cite{32}.
	\item $\bm{C}_t\in\mathbb{R}_+^{30}$: Commodity Channel Index (CCI) is calculated using high, low and close price. CCI compares current price to average price over a time window to indicate a buying or selling action\cite{33}.
	\item $\bm{X}_t\in\mathbb{R}^{30}$: Average Directional Index (ADX) is calculated using high, low and close price of each stock at current time step t. ADX identifies trend strength by quantifying the amount of price movement\cite{34}.
\end{enumerate}

\subsubsection{Action Space}
A set containing $2k+1$ elements represents the action space of the multi-stock trading environment: $\{-k,...,-1,0,1,...,k\}$, where $k,-k$ represents the number of shares we can buy and sell at once. It satisfies the following conditions: 

\begin{enumerate}
	\item 	$h_{max}$ represents the maximum number of shares we can be able to buy at a time.
	\item The action space is a high-dimensional and large discrete action space since the entire action space is of size $\left(2k+1\right)^{30}$, and it can be approximately considered as a continuous action space in the practice.
	\item The action space will next be normalized to $[-1,1]$.
\end{enumerate}

\subsubsection{Reward}
We define the reward value of the multi-stock trading environment as the change in portfolio value from state $s$ taking action a to the next state $s^\prime$ (in this case two days before and after), with the training objective of obtaining a trading strategy that maximizes the return:
\begin{equation}
	Return_t\left(s_t,\ a_t,\ s_{t+1}\right)=\left(b_{t+1}+\bm{p}_{t+1}^T\bm{h}_{t+1}\right)-\ \left(b_t+\bm{p}_t^T\bm{h}_t\right)-c_t
\end{equation}
where $c_t$ represents the transaction cost. We assume that the per-transaction cost is 0.1\% of each transaction, as defined in Yang\cite{10}:
\begin{equation}
	c_t=0.1\%\ \cdot\ \left |\bm{p}^T\bm{k}_t  \right | 
\end{equation}

\subsubsection{Turbulence Threshold}
We employ this financial index $turbulence_t$\cite{14} that measures extreme asset price movements to avoid the risk of sudden events that may cause stock market crash\cite{35}, such as March 2020 stock market caused by COVID-19, wars and financial crisis:
\begin{equation}
	turbulence_t=\left(\bm{y}_t-\bm{\mu}\right)\bm{\Sigma}^{-1}\left(\bm{y}_t-\bm{\mu}\right)^\prime\in\mathbb{R}
\end{equation}
Where $\bm{y}_t\in\mathbb{R}^{30}$ denotes the stock returns for current period t, $\bm{\mu}_t\in\mathbb{R}^{30}$ denotes the average of historical returns, and $\bm{\Sigma}\in\mathbb{R}^{30\times30}$ denotes the covariance of historical returns. 
Considering the historical volatility of the stock market, we set the turbulence threshold to 90th percentile of all historical turbulence indexes. If $turbulence_t$ is greater than this threshold, it means that extreme market conditions are occurring and the agent will stop trading until the turbulence index falls below this threshold.

\subsubsection{Other Parameters}
In addition to defining the state space, action space and reward functions, some necessary constraints need to be added to the multi-stock trading environment.
\begin{enumerate}
	\item Initial capital: \$1 million.
	\item Maximum number of shares in a single trade $h_{max}$: 100.
	\item Reward scaling factor: 1e-4, which means the reward returned by the environment will be only 1e-4 of the original one.
\end{enumerate}
	
\subsection{Stock Trading Agent}
\subsubsection{Framework}
We introduce LSTM as a feature extractor to improve the model in Yang\cite{14}, as shown in Figure 2. We refer to the PPO algorithm that incorporates LSTM as a feature extractor as the PPO with Cascaded LSTM (\textbf{CLSTM-PPO}) model because we also use the LSTM policy in PPO.
\begin{figure}[htbp!]
	\centering
	\includegraphics[width=1.0\linewidth]{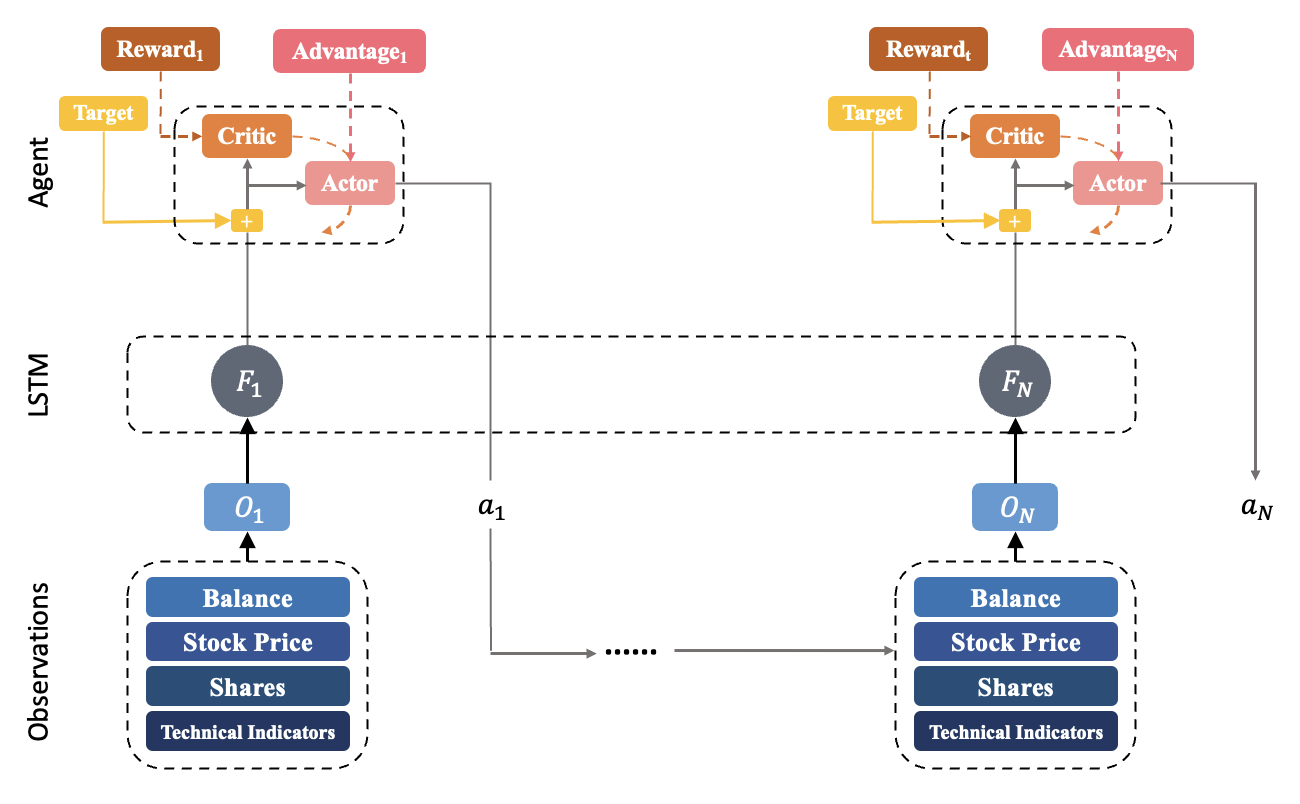}
	\caption{Overview of our CLSTM-PPO model}
	\label{overview}
\end{figure}

At the time step $t$, the environment automatically generates the current state $S_t$ and passes it to the LSTM network. It remembers the state and uses its memory to retrieve the past $T$ stock market states to obtain the state sequence $F_t=\ [S_{t-T+1},...,S_t]$. The LSTM analyzes and extracts the hidden time-series features or potentially profitable patterns in $F_t$, and then outputs the encoded feature vector $F_t^\prime$ and passes it to the agent, which is guided by the policy function $\pi(F_t^\prime)$ to perform the action $a_t$ that can get a good return. The environment then returns the reward $R_t$ , the next state $S_{t+1}$ , and a boolean $d_t$ to determine if the state is terminated according to the agent's behavior. Then, the obtained quintet $(S_t,\ a_t,\ R_t,\ S_{t+1},\ d_t)$ is stored in the experience pool. Actor computes $A_t$ from the $targets_t$ computed by critic using the advantage function. After a certain number of steps, actor back-propagates the error through the clipped surrogate objective function of PPO, then critic updates the parameters using the mean square error loss function. The environment will keep repeating the process until the end of the training phase.

\subsubsection{LSTM as Feature Extractor}
Reinforcement Learning (RL) was initially applied to games, which have a limited action space, clear stopping conditions and a more stable environment, so there is room to improve the use of RL for stock trading. It is well known that financial markets are full of noise and uncertainty, and that the factors affecting stock prices are multiple and changing over time. This makes the stock trading process more like a partially observable Markov decision process (POMDP), since the states we use are not the real states in the stock trading environment. Therefore, we can use the memory property of the LSTM to discover features of the stock market that change over time. LSTM can integrate information hidden in the time dimension, thus making it more likely that the POMDP is closer to the MDP\cite{24}\cite{31}.  

In this paper, an LSTM-based feature extractor is developed using the customized feature extraction interface provided by stable-baselines3\footnote{Github repository: https://github.com/DLR-RM/stable-baselines3}, which is a set of reliable implementations of reinforcement learning algorithms in PyTorch. The network structure of the LSTM feature extractor is shown in Figure 3.

\begin{figure}[htbp!]
	\centering
	\includegraphics[width=1.0\linewidth]{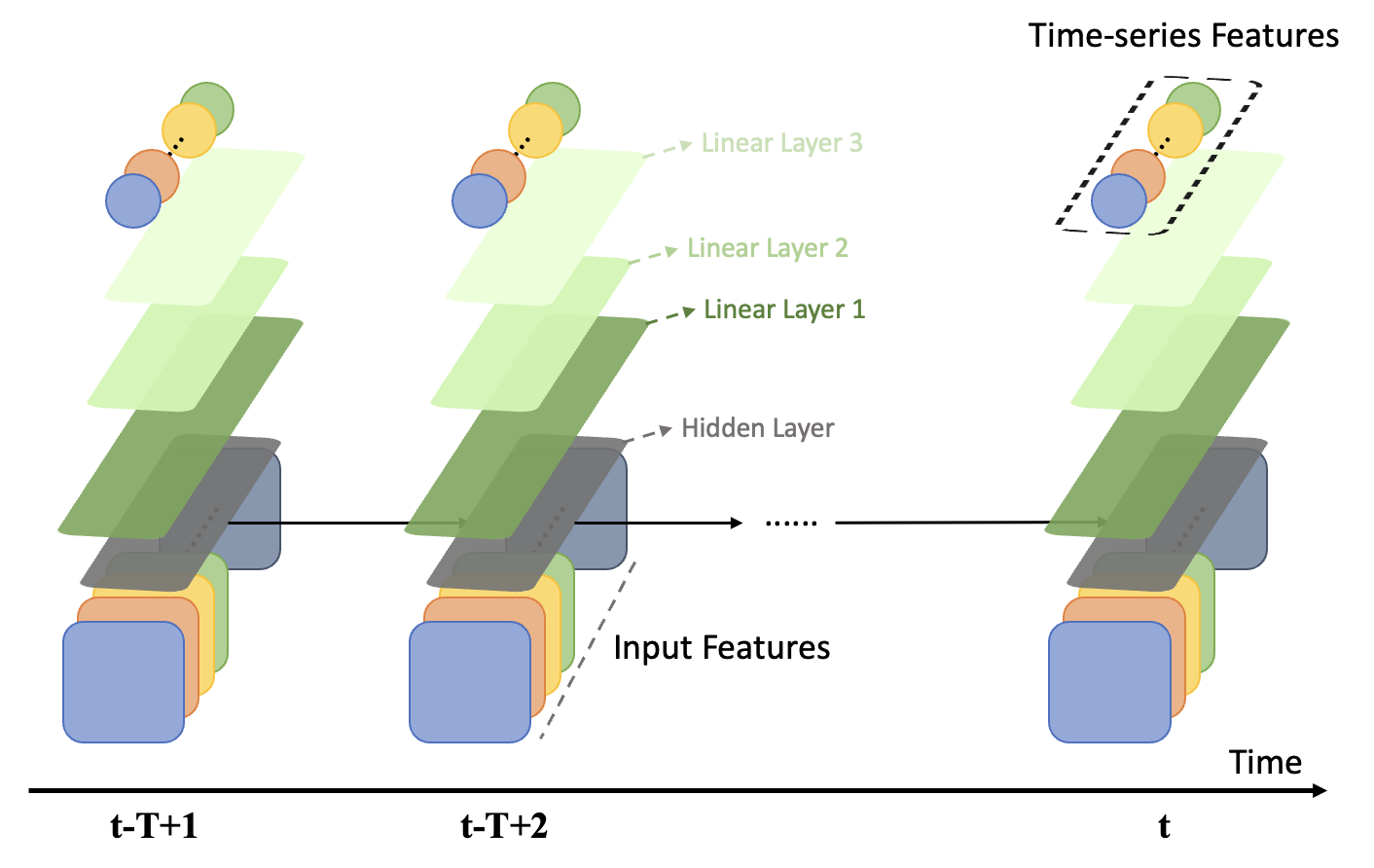}
	\caption{Overview of our model}
	\label{lstm}
\end{figure}

As shown in the Figure 3, each input to the LSTM is a state list of length T arranged in time order. Starting from the farthest state, the hidden layer of the LSTM remembers the information of the state and passes it to the next point in time. We use the feature vector of the most recent state, which is the current state, after one hidden layer of the LSTM with three linear layers. This feature vector $F_t^\prime$ will be used as input feature, then be put into the PPO for the agent to learn. 

Algorithm 1 is the pseudo-code of LSTM as a feature extractor. In practice, the defined LSTM can be connected to the reinforcement learning algorithm using the policy\_kwargs interface provided by stable-baselines3, which enables the agent to directly receive the temporal features extracted by the LSTM.

\begin{algorithm}[htbp!]
    \caption{one-day LSTM feature extractor}
    \begin{small}
    \BlankLine
    \KwIn{
    hidden state of shape $h_0=(num\_layers*num\_directions, N, hidden\_size)$,
    cell state of shape $c_0=(num\_layers*num\_directions, N, hidden\_size)$}
    \KwOut{$N$-day time-series feature}
    Get last $N$-day states list\;
    Initialize LSTM hidden and cell states: $h = h_0$, $c = c_0$\;
    \For{n in range($N$)}{
        Pass $n$th state into LSTM\;
        Store the output and update the LSTM with $(h,c)$ in output\;
    }
    Extract features from the last LSTM layer\;
    Return features
    \end{small}
\end{algorithm}

\subsubsection{Proximal Policy Optimization (PPO)}
PPO\cite{37} is one of the most advanced of the current policy-based approaches which use multiple epochs of stochastic gradient ascent to perform each policy update\cite{39}, and also performs best in stock trading among the three DRL algorithms in Yang\cite{14}, which is an important reason for our consideration of it. In PPO, parameters of actor (agent) is $\theta$. 

First, a notation is used to represent the probability ratio between the new policy and the old one:
\begin{equation}
	r_t\left(\theta\right)=\ \frac{\pi_\theta(a_t|s_t)}{\pi_{\theta_{old}}(a_t|s_t)}
\end{equation}
So we can get $r_t\left(\theta_{old}\right)=1$ from it.
The clipped surrogate objective function of PPO is:
\begin{equation}
	\begin{aligned}
		H^{CLIP}\left(\theta\right)&=\hat{\mathbb{E}}[\min(r_t\left(\theta\right)\hat{A}\left(s_t,\ a_t\right),\\ &clip\left(r_t\left(\theta\right),r_t\left(\theta_{old}\right)-\epsilon,r_t\left(\theta_{old}\right)+\epsilon\ \right)\hat{A}\left(s_t,\ a_t\right))]
	\end{aligned}
\end{equation}
That is,
	\begin{equation}
		\begin{aligned}
		  H^{CLIP}\left(\theta\right)&=\hat{\mathbb{E}}[\min(r_t\left(\theta\right)\hat{A}\left(s_t,\ 	a_t\right),\\ &clip\left(r_t\left(\theta\right),1-\epsilon,1+\epsilon\ \right)\hat{A}\left(s_t,\ a_t\right))]
		\end{aligned}
	\end{equation}
Where $r_t\left(\theta\right)\hat{A}\left(s_t,\ a_t\right)$ is the normal policy gradient objective, and $\hat{A}\left(s_t,\ a_t\right)$ is the estimated advantage function. Term clip$\left(r_t\left(\theta\right),1-\epsilon,1+\epsilon\ \right)$ clips the ratio $r_t\left(\theta\right)$ to be within $[1-\epsilon,1+\epsilon]$. Finally, $H^{CLIP}(\theta)$ takes the minimum of the clipped and unclipped objective.


The operation of LSTM in PPO is similar to the LSTM in above section, which is equivalent to making the agent recall the behavioral information of the previous moment while receiving new data, so that the decision made by the agent at this moment is based on the previous decision.

In the LSTM network, the initial number of features is 181, the final number of output features is 128 and the hidden size is 128. Linear layer 1 is $(15\times128,\ 128)$ and then passes the Tanh activation function. Linear layer2 and 3 are the same two layers of $(128,\ 128)$, and  then pass Tanh.

Algorithm 2 is our pseudo-code for training agent using Proximal Policy Optimization (PPO) combined with cascaded Long Short-Term Memory (LSTM) Network.

\begin{algorithm}[h]
\SetAlgoLined
\KwIn{Initial state $s_t$ ;
Adam optimizer with learning rate $\alpha$;
Discount factor $\gamma$;
Clipping range $\epsilon$;
Advantage estimate $A_t$;}
\KwOut{Trained actor network $\pi_{\theta}(a_t|s_t)$ and value network $V_{\phi}(s_t)$;}
Initialize critic $V_\phi(s)$ and actor $\pi_\theta(a|s)$ networks with parameters $\phi$ and $\theta$\;
Initialize the replay buffer $D$\;
\For{each episode}{
    Initialize the environment with initial state $s_0$\;
    \For{each step $t$ in the episode}{
        Receive state $s_t$ from environment\;
        Process $s_t$ with LSTM to obtain a feature vector $f_t$\;
        Compute the critic's value estimate $\hat{v}_t=V_\phi(f_t)$\;
        Sample an action $a_t$ from the policy $\pi_\theta(a_t|f_t)$\;
        Execute $a_t$ in the environment to receive the reward $r_t$ and the next state $s_{t+1}$\;
        Compute the advantage estimate $A_t = r_t + \gamma \hat{v}_{t+1} - \hat{v}_t$\;
        Add the transition $(f_t, a_t, A_t)$ to the replay buffer $D$\;
        \If{$t \mod T = 0$}{
            Update the critic by minimizing the MSE between the target $r_t + \gamma \hat{v}_{t+1}$ and the current estimate $\hat{v}_t$: $\phi \leftarrow \phi - \alpha_V \nabla_\phi (r_t + \gamma \hat{v}_{t+1} - \hat{v}_t)^2$\;
            Update the actor using the PPO objective function: $\theta \leftarrow \theta + \alpha_\theta \nabla_\theta L_{\rm PPO}(\theta)$\;
            Clear the replay buffer $D$\;
        }
    }
}
\caption{PPO with LSTM}
\end{algorithm}

\section{Performance Evaluations}
In this section, we first tuned some parameters in the model, then used 30 Dow constituent stocks to evaluate our model, and performed robustness tests on 30 stocks in the SSE 50 in China, 30 stocks in SENSEX in India, and 30 stocks in FTSE 100 in UK.

\subsection{Description of Datasets}
In this paper, 120 stocks are selected as the stock pool: 30 Dow constituent stocks, 30 stocks randomly selected from the SSE50 which is an index of the top 50 companies listed on the Shanghai Stock Exchange in Chinese stock market, 30 stocks from SENSEX which is composed of 30 of the largest and most actively traded stocks on the Bombay Stock Exchange (BSE), representing various sectors of the Indian economy and 30 stocks from FTSE 100 which consists of the 100 largest and most highly capitalized companies listed on the London Stock Exchange (LSE) by market capitalization. The stocks from Dow Jones are the same as the pool in Yang[14] to facilitate comparison with their ensemble strategy, while the 90 stocks from FTSE 100, SSE 50, and SENSEX are used to explore the applicability of this paper's model in a mature capital market as well as two emerging markets.

The daily data for backtesting starts from 01/01/2009 and ends on 05/08/2020, and the data set is divided into two parts: in-sample period and out-sample period. The data in the in-sample period is used for training, and the data in the out-sample period is used for trading. We only use PPO during the whole process.

The entire dataset is split as shown in the Figure 4. The training data is from 01/01/2009 to 12/31/2015, and the trading data is from 01/01/2016 to 05/08/2020. In order to better exploit the data and allow the agent to better adapt to the dynamic changes of the stock market, the agent can continue to be trained during the trading phase. Using data from the U.S. market as an example, the first training was started from 01/01/2009 until 01/01/2016, and then traded on the test set for three months. The second training will start from 01/01/2009 until 03/01/2016, then trade from 03/02/2016 until 06/01/2016. and so on until the last quarter of the test set.

It should be noted that in both markets, India and UK, it is not possible to ensure that all stocks are traded from 01/01/2009 because of the changes in the constituent stocks of the index, so after data pre-processing, there are a total of 7 years of data for the Indian market and close to 5 years of data for the UK market. The Indian market is trained from 02/26/2016 until 03/02/2022 and traded from 03/03/2022 until 03/03/2033. the UK market is trained from 09/19/2018 until 03/02/2022 and traded from 03/03/2022 until 03/03/2033.

\begin{figure}[htbp!]
	\centering
	\includegraphics[width=1.0\linewidth]{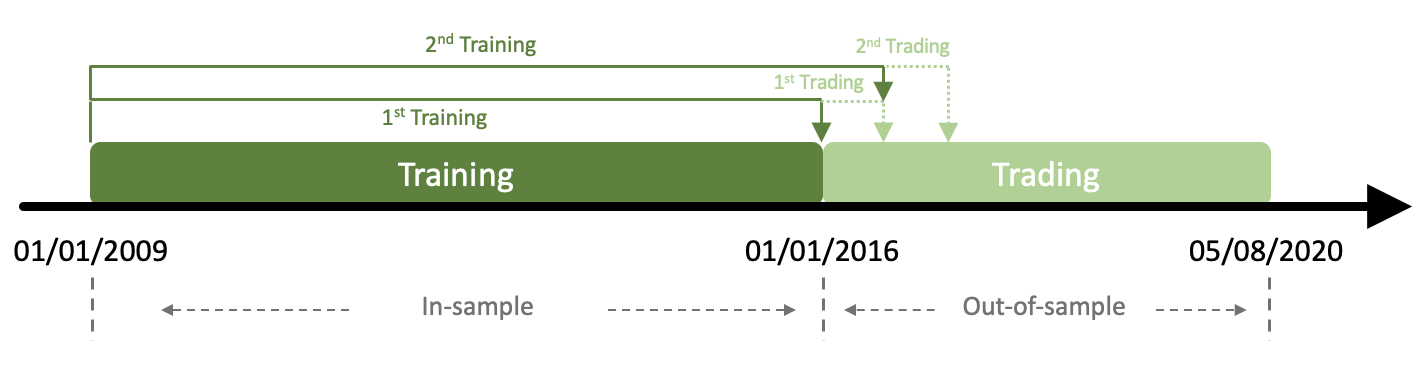}
	\caption{Stock Data Splitting}
	\label{datasplit}
\end{figure}

\subsection{Training Parameters of PPO}
The agent is trained using only the actor-critic based PPO method, and the training parameters of PPO are set as shown in the Table 1. 
\begin{table}[htbp!]
	\centering
	\caption{Training parameters of PPO}
	\begin{tabular}{l|lll} 
		\hhline{==~~}
		\textbf{Parameter}                           & \textbf{Value} &  &   \\ 
		\cline{1-2}
		Reward Discount Factor                       & 0.99           &  &   \\ 
		\cline{1-2}
		Update Frequency                             & 128            &  &   \\ 
		\cline{1-2}
		Loss Function Weight of Critic               & 0.5            &  &   \\ 
		\cline{1-2}
		Loss Function Weight of Distribution Entropy & 0.01           &  &   \\ 
		\cline{1-2}
		Clip Range                                   & 0.2            &  &   \\ 
		\cline{1-2}
		Maximum of Gradient Truncation               & 0.5            &  &   \\ 
		\cline{1-2}
		Optimizer                                    & Adam           &  &   \\ 
		\cline{1-2}
		$\beta_1$                      & 0.9            &  &   \\ 
		\cline{1-2}
		$\beta_2$                      & 0.999          &  &   \\ 
		\cline{1-2}
		$\epsilon$                      & 1e-8           &  &   \\ 
		\cline{1-2}
		Learning Rate                                & 3e-4           &  &   \\
		\hhline{==~~}
	\end{tabular}
\end{table}

\subsection{Baseline Methods}
Our model is compared with baseline models including:
\begin{itemize}
	\item \textbf{Buy and hold Dow Jones Index}: the typical Buy-And-Hold strategy, which means that the trader buys at the start of the trading period and holds to the end.
	\item \textbf{Buy and hold SSE50 Index}: a Buy-And-Hold strategy in Chinese market.
        \item \textbf{Buy and hold SENSEX Index}: a Buy-And-Hold strategy in Indian market.
        \item \textbf{Buy and hold FTSE 100 Index}: a Buy-And-Hold strategy in UK market.
        \item \textbf{PPO model}: only use PPO with MLP Policy to train the agent.
        \item \textbf{Recurrent PPO model}: use PPO with LSTM Policy to train the agent.
        \item \textbf{MLP model}: a model offered by Qlib\cite{51}, which can predict the stock price using Multilayer Perceptron (MLP).
        \item \textbf{LSTM model}: a model offered by Qlib\cite{51}, which can use Long Short-Term Memory networks to predict the stock price.
        \item \textbf{Light GBM model}: Light Gradient Boosting Machine is a highly efficient gradient boosting framework that uses a novel decision tree algorithm based on histogram-based computation. It is designed to handle large-scale datasets and provides faster training speed and higher accuracy than other gradient boosting models.
        \item \textbf{HIST model}: Histogram-based Gradient Boosting is another gradient boosting framework that is based on histogram-based algorithm. It is designed for high-dimensional and sparse data, and is able to handle datasets with a large number of features. HIST provides better performance than traditional gradient boosting models when dealing with high-dimensional data.
	\item \textbf{Ensemble Strategy in Yang\cite{14}}: they train agents for three months simultaneously in the training phase using A2C, DDPG and PPO algorithms and then select the agent with the highest Sharpe ratio as the trader for the next quarter. This process is repeated until the end of the training.
\end{itemize}

\subsection{Evaluation Measures}
\begin{itemize}
	\item \textbf{Cumulative Return (CR)}: calculated by subtracting the portfolio’s final value from its initial value, and then dividing by the initial value. It reflects the total return of a portfolio at the end of trading stage.
	\begin{equation}
		CR=\frac{P_{end}-P_0}{P_0}
	\end{equation}
	\item \textbf{Max Earning Rate (MER)}: the maximum percentage profit during the trading period. It measures the robustness of a model and reflects the trader’s ability to discover the potential maximum profit margin.
	\begin{equation}
		MER=\frac{max(A_x-\ A_y)}{A_y}\ 
	\end{equation}
	Where $A_x,\ A_y$ is the total asset of the strategy and $x>y,\ A_y<A_x$.
	\item \textbf{Maximum Pullback (MPB)}: the maximum percentage loss during the trading period. It measures the robustness of a model.
	\begin{equation}
		MPB=\ \frac{max(A_x-\ A_y)}{A_y}
	\end{equation}
	Where $A_x,\ A_y$ is the total asset of the strategy and $x>y,\ A_y>A_x$.
	\item \textbf{Average Profitability Per Trade (APPT)}: refers to the average amount that you can expect to win or lose per trade. It measures the trading performance of the model.
	\begin{equation}
		APPT=\frac{P_{end}-P_0}{NT}
	\end{equation}
	Where $P_{end}-P_0$ means the returns at the end of trading stage, and $NT$ is the number of trades. 
	\item \textbf{Sharpe Ratio (SR)}: calculated by subtracting the annualized risk free rate from the annualized return, and the dividing by the annualized volatility. It considers benefits and risks synthetically and reflects the excess return over unit systematic risk.
	\begin{equation}
		SR=\ \frac{E\left(R_P\right)-R_f}{\sigma_P}
	\end{equation}
\end{itemize}

\subsection{Exploration of Well-performing Hyperparameters}
We performed parameter tuning on two important parts of the model: (i) the time window size of the LSTM as a feature extractor and (ii) the hidden size of LSTM in PPO training.
\subsubsection{Best Time Window of LSTM}
For the time window of the LSTM, we tested the cases of TW=5,15,30,50, and then show the trading results of the model in Figure 5 (hidden size of  LSTM  in PPO is 512).
\begin{figure*}[htbp!]
	\centering
	\includegraphics[width=1.0\linewidth]{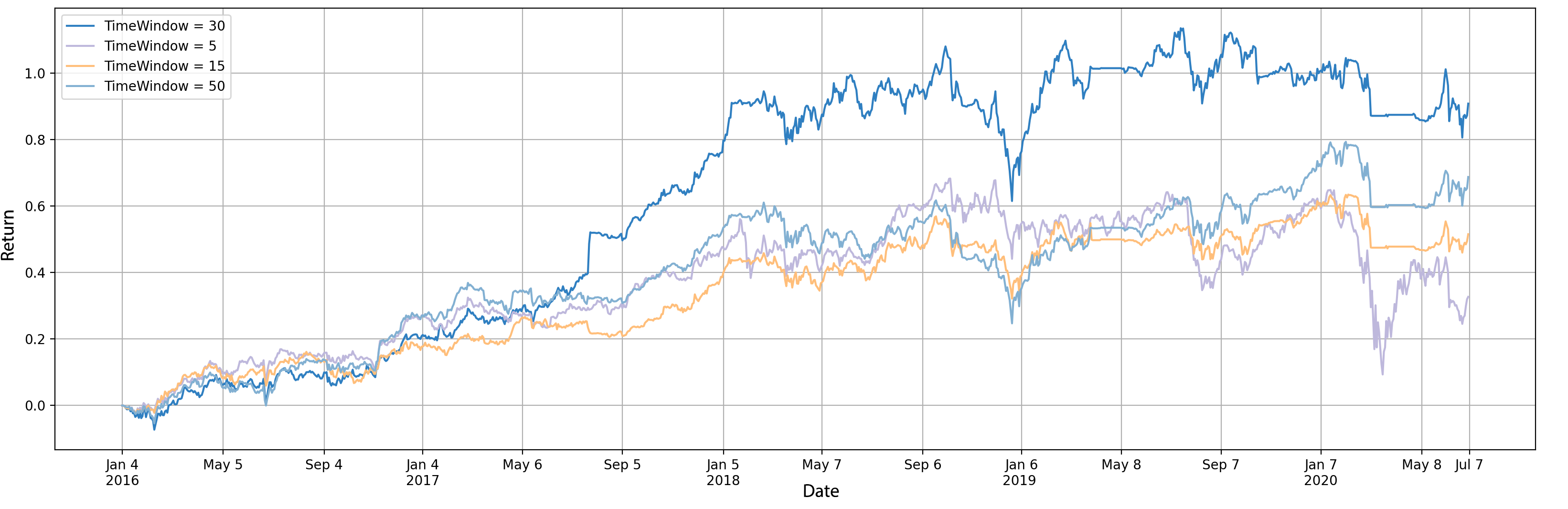}
	\caption{Trading results of different time windows in LSTM}
	\label{para_lstm}
\end{figure*}
From the Figure 5, it can be seen that the agent at TW=30 is able to achieve the highest cumulative return during the trading period, ahead of the agent at TW=50 by more than 20\% and a figure that exceeds 40\% in agent without LSTM extraction. This verifies the feasibility of our model: the stock price movements in the stock market are correlated with their past trajectories, and the LSTM is able to extract the time-series features of them.
\begin{table}[htbp!]
	\centering
	\arrayrulecolor{black}
	\caption{Comparison of different time windows in LSTM}
	\begin{tabular}{lllll} 
		\hline
		~    & TW=5    & TW=15            & TW=30             & TW=50    \\ 
		\hline
		CR   & 32.69\% & 51.53\%          & \textbf{90.81\%}  & 68.74\%  \\
		MER  & 68.27\% & 63.46\%          & \textbf{113.50\%} & 79.32\%  \\
		MPB  & 58.93\% & \textbf{24.75\%} & 46.51\%           & 37.01\%  \\
		APPT & 18.29   & 21.77            & \textbf{35.27}    & 23.31    \\
		SR   & 0.2219  & 0.7136           & \textbf{1.1540}   & 0.9123   \\
		\hline
	\end{tabular}
	\arrayrulecolor{black}
\end{table}

The data in the Table 2 shows the difference between these options in more detail: for TimeWindow=30, CR, MER, APPT and SR are much higher than the other options, but MPB does not perform well. On balance, TW = 30 is the best parameter for our experiment.

\subsubsection{Best Hidden Size of LSTM in PPO}
For the hidden size of the LSTM  in PPO, we tested the cases of HS=128, 256, 512, 1024, 512*2(two hidden layers) and then show the trading results of the model in Figure 6. (time window of LSTM is 30)
\begin{figure*}[ht]
	\centering
	\includegraphics[width=1.0\linewidth]{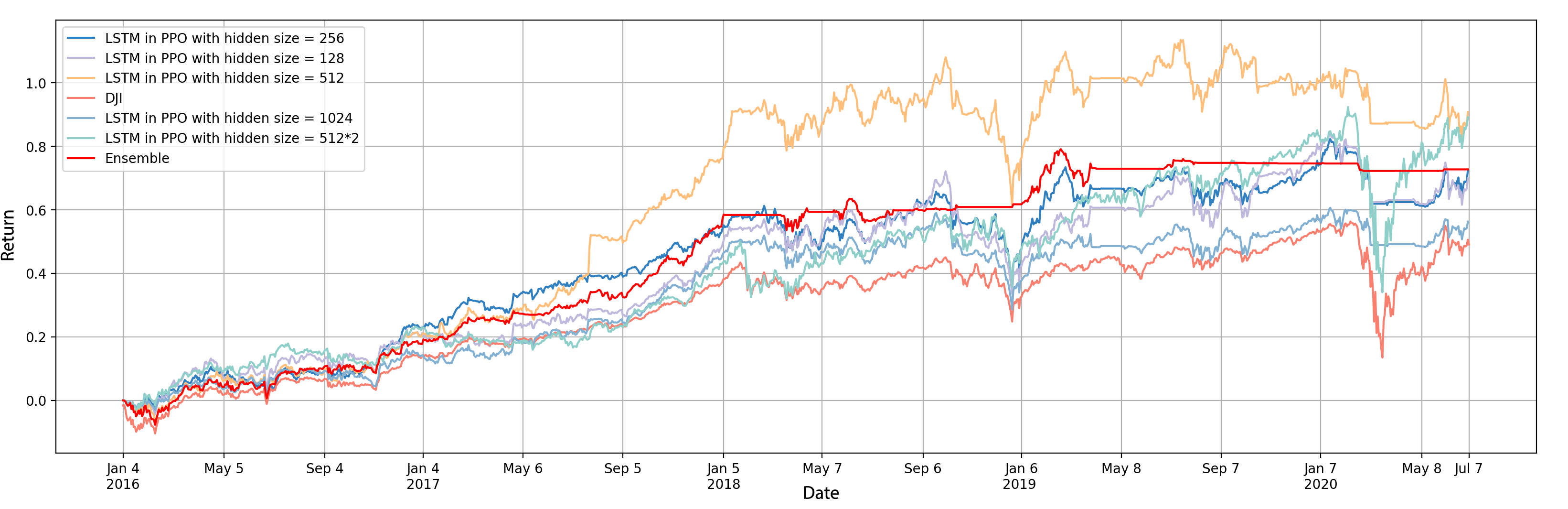}
	\caption{Trading results of different hidden sizes of LSTM in PPO}
	\label{para_ppo}
\end{figure*}

As can be seen from the Figure 6, when hidden size=512, the cumulative yield is significantly higher than the other choices. It has a smaller drawback compared to hidden size=512*2 and was able to stop trading in the big drawback in March 2020, indicating that the agent can be a smart trader under the right training conditions of DRL.
\begin{table}[htbp!]
	\centering
	\caption{Comparison of different hidden sizes of LSTM in PPO}
	\arrayrulecolor{black}
	\resizebox{\linewidth}{!}{
	\begin{tabular}{llllll} 
		\hline
		~    & HS=128  & HS=256           & HS=512            & HS=1024 & HS=512*2  \\ 
		\hline
		CR   & 69.94\% & 72.58\%          & \textbf{90.81\%}  & 56.27\% & 89.13\%   \\
		MER  & 84.29\% & 82.32\%          & \textbf{113.50\%} & 60.64\% & 92.34\%   \\
		MPB  & 38.57\% & \textbf{29.92\%} & 46.51\%           & 30.39\% & 58.31\%   \\
		APPT & 28.07   & 30.79            & \textbf{35.27}    & 23.96   & 33.26     \\
		SR   & 0.9255  & 1.0335           & \textbf{1.1540}   & 0.8528  & 0.8447    \\
		\hline
	\end{tabular}}
	\arrayrulecolor{black}
\end{table}

The data in the Table 3 shows the difference between these options in more detail: for HiddenSize=512, CR, MER, APPT and SR are much higher than the other options, but MPB does not perform well. On balance, HS = 512 is the best parameter for our strategy.

\subsection{Performance in U.S. Markets}
The well-performing parameter set (TW=30, HS=512) is used as the parameters of the final model and the results of this model are compared with the trading results of the PPO model with LSTM in PPO and another with MlpPolicy and the Ensemble Strategy in Yang\cite{14}, as shown in Figure 7.

\begin{figure*}[ht]
	\centering
	\includegraphics[width=1.0\linewidth]{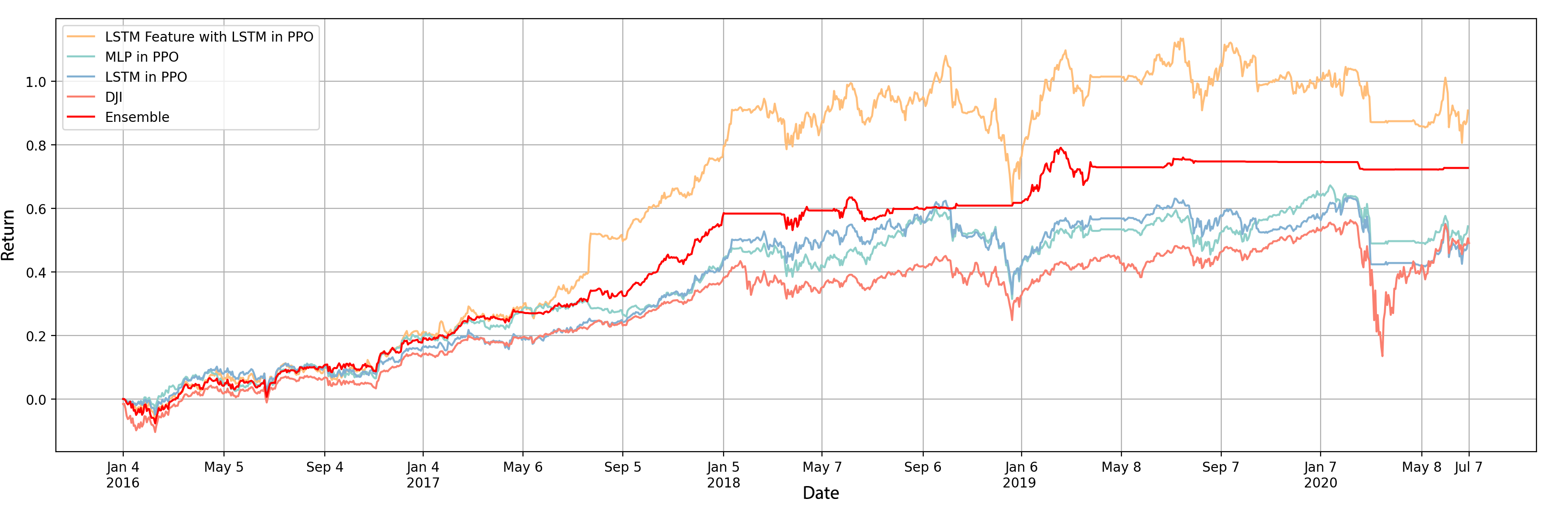}
	\caption{Trading results by agents with LSTM in PPO plus LSTM feature extraction, LSTM in PPO, Ordinary PPO, Ensemble Strategy in Yang\cite{14} and Buy-And-Hold strategy on DJI}
	\label{usmarket}
\end{figure*}

Table 4 shows the details of trading results in U.S. Markets.
\begin{table}[htbp!]
	\centering
	\arrayrulecolor{black}
	\caption{Details of the trading results in U.S. Markets}
	\scalebox{0.71}{
	\begin{tabular}{lllllll} 
		\cline{1-6}
		~    & PPO     & RecurrentPPO & CLSTM-PPO         & Ensemble         & DJI     & ~  \\ 
		\cline{1-6}
		CR   & 54.37\% & 49.77\%     & \textbf{90.81\%}  & 70.40\%          & 50.97\% & ~  \\
		MER  & 67.28\% & 63.45\%     & \textbf{113.50\%} & 65.32\%          & 63.90\% & ~  \\
		MPB  & 28.30\% & 29.39\%     & 46.51\%           & \textbf{15.74\%} & 72.32\% & ~  \\
		APPT & 20.02   & 22.84       & \textbf{35.27}    & 28.54            & N.A.    & ~  \\
		SR   & 0.8081  & 0.6819      & 1.1540            & \textbf{1.3000}  & 0.4149  & ~  \\
		\cline{1-6}
	\end{tabular}}
	\arrayrulecolor{black}
\end{table}

Our model (CLSTM-PPO: PPO with cascaded LSTM networks) obtains the highest cumulative return of 90.81\% and the maximum profitability of 113.5\% on 30 Dow components, better than the ensemble strategy in Yang[14], the baseline model. However, in terms of maximum pullback, our model has an MPB=45.51\% compared to DJI's MPB=63.90\%, indicating the possession of risk tolerance and the ability to identify down markets and stop trading. In this respect, ensemble strategy does a little better. But for a long time the agent of ensemble strategy has a negative attitude towards investing, choosing not to trade whether the market is down or up. This can lead to the loss of a large profit margin in the long run. Overall, our model has the strongest profit-taking ability, excels at finding profits within volatile markets, and recovers quickly after pullbacks. In terms of the Sharpe ratio, our strategy is very close to the baseline model, but does not require the help of other algorithms, which is much easier and faster.

Analyzing in more detail, all strategies can be divided into two phases which are consistent with DJI index: (i) Accumulation phase: until 06/2017, our strategies were able to achieve stable growth with little difference in returns from the integrated strategies. However, after that, our agent quickly captured profits and was able to grow total returns rapidly. This phase lasted until 01/2018, when the cumulative return had reached a level where the difference with the final return was not significant. (ii) Volatility phase: Starting from 01/2018, our agent's trading style became very aggressive and courageous, as reflected in the large fluctuations in returns. The returns were generally more stable during this phase and were able to bounce back quickly within two months after suffering a pullback on 01/2019.

\subsection{Performance in Chinese Markets}
The same model is used to trade the samples in the Chinese stock market, as Figure 8 shown, and the ensemble strategy is chosen as the most important baseline model. We show the cumulative returns at the end of the trade as follows.
\begin{figure*}[ht]
	\centering
	\includegraphics[width=1.0\linewidth]{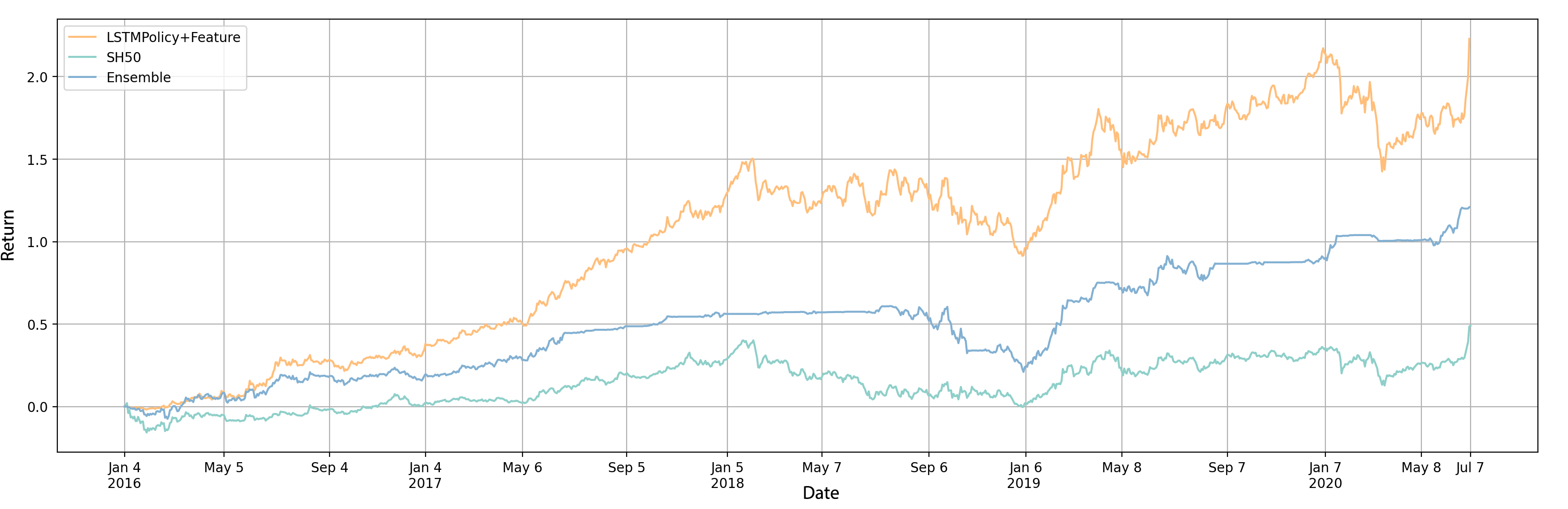}
	\caption{Trading results by agents with LSTM in PPO plus LSTM feature extraction, Ensemble Strategy in Yang[14] and Buy-And-Hold strategy on SSE50}
	\label{cnmarket}
\end{figure*}

Table 5 shows the details of trading results in Chinese Markets.
\begin{table}[htbp!]
	\centering
	\arrayrulecolor{black}
	\caption{Details of the trading results in Chinese Markets}
	\scalebox{0.71}{\begin{tabular}{llllll} 
		\hline
		~ & PPO & RecurrentPPO & CLSTM-PPO         & Ensemble         & SSE50    \\ 
		\hline
		CR   & 93.23\% & 102.48\% &\textbf{222.91\%} & 120.87\%         & 51.46\%  \\
		MER  & 93.23\%&102.48\%&\textbf{222.91\%} & 120.87\%         & 51.46\%  \\
		MPB  & 32.48\%&33.92\%&74.81\%           & \textbf{39.95\%} & 41.27\%  \\
		APPT & 39.44&42.38&\textbf{66.96}    & 47.55            & 25.78    \\
		SR   & 1.5489&1.5977&\textbf{2.3273}   & 1.6938           & 0.4149   \\
		\hline
	\end{tabular}}
	\arrayrulecolor{black}
\end{table}

In the Chinese market, the advantages of our model (CLSTM-PPO: PPO with cascaded LSTM networks) are greater than the ensemble strategy. Our CR=222.91\% is nearly twice that of ensemble strategy (Ensemble\_CR=120.87\%). Although our pullbacks are greater, the volatility is more reflected in the rise in cumulative returns. Also, the Sharpe ratio tells us that our model (SR=2.3273) is better in the Chinese market when combining return and risk. This illustrates our model's superior performance in emerging markets, which typically have greater volatility, and is consistent with our analysis of our model: the ability to capture returns in volatility quickly and accurately, and the returns obtained are positively correlated with positive volatility.

On further analysis, the cumulative return of our model increased rapidly from 01/2016 to 02/2018, finally reaching almost three times that of the integrated strategy. Subsequently, as the volatility of the SSE 50 increased, the volatility of our model increased accordingly. Within 02/2018 to 01/2019, the pullbacks of the two models were similar, but then our model captured nearly 80\% of the return at 01/2019 within three months. Even though it suffered a decline in 01/2020 due to the black swan event of Covid-19, it quickly bounced back to its highest point six months later.

\subsection{Performance in Indian and UK markets}
In this section, to fully test the robustness of our CLSTM-PPO model, we have also tested it in the UK market and the Indian market, while also using the state-of-the-art trading models provided by Qlib\cite{51} as a benchmark model. Finally, we put together the trading data of our model and all benchmark models in the four markets as a comparison, as shown in Table 6.
\begin{table*}[ht]
\centering
\caption{Comparison of experimental results with baselines}
\resizebox{\textwidth}{!}{\begin{tabular}{|ccccccccccc|}
\hline
\textbf{Datasets}                                       & \textbf{Metrics}                   & \textbf{PPO} & \textbf{RecurrentPPO} & \textbf{Ensemble}                       & \textbf{Buy Index}         & \textbf{MLP} & \textbf{LSTM} & \textbf{lightGBM}                       & \textbf{HIST}                           & \textbf{CLSTM-PPO}                       \\ \hline
\multicolumn{1}{|c|}{}                                  & \multicolumn{1}{c|}{\textbf{CR}}   & 54.37\%      & 49.77\%               & 70.40\%                                 & 50.97\%                                & 51.27\%      & 45.55\%       & 36.68\%                                 & 87.27\%                                 & { \textbf{90.81\%}}  \\
\multicolumn{1}{|c|}{}                                  & \multicolumn{1}{c|}{\textbf{MER}}  & 67.28\%      & 63.45\%               & 65.32\%                                 & 63.90\%                                & 65.77\%      & 61.18\%       & 47.67\%                                 & 94.33\%                                 & {\textbf{113.50\%}} \\
\multicolumn{1}{|c|}{}                                  & \multicolumn{1}{c|}{\textbf{MPB}}  & 28.30\%      & 29.39\%               & { \textbf{15.74\%}} & 72.32\%                                & 29.19\%      & 25.97\%       & 18.33\%                                 & 20.72\%                                 & 46.51\%                                  \\
\multicolumn{1}{|c|}{}                                  & \multicolumn{1}{c|}{\textbf{APPT}} & 20.02        & 22.84                 & 28.54                                   & N.A.                                   & 16.02        & 23.06         & 17.48                                   & 33.22                                   & { \textbf{35.27}}    \\
\multicolumn{1}{|c|}{\multirow{-5}{*}{\textbf{USA}}}    & \multicolumn{1}{c|}{\textbf{SR}}   & 0.8081       & 0.6819                & 1.3116                                  & 0.4149                                 & 0.4368       & 0.8471        & 0.7787                                  & { \textbf{1.4884}}  & 1.154                                    \\ \hline
\multicolumn{1}{|c|}{}                                  & \multicolumn{1}{c|}{\textbf{CR}}   & 93.23\%      & 102.48\%              & 120.87\%                                & 51.46\%                                & 54.32\%      & 79.93\%       & 39.62\%                                 & 147.57\%                                & { \textbf{222.91\%}} \\
\multicolumn{1}{|c|}{}                                  & \multicolumn{1}{c|}{\textbf{MER}}  & 93.23\%      & 102.48\%              & 120.87\%                                & 51.46\%                                & 54.32\%      & 79.93\%       & 39.62\%                                 & 147.57\%                                & { \textbf{222.91\%}} \\
\multicolumn{1}{|c|}{}                                  & \multicolumn{1}{c|}{\textbf{MPB}}  & 32.48\%      & 33.92\%               & 39.95\%                                 & 41.27\%                                & 25.56\%      & 23.12\%       & { \textbf{15.83\%}} & 29.86\%                                 & 74.81\%                                  \\
\multicolumn{1}{|c|}{}                                  & \multicolumn{1}{c|}{\textbf{APPT}} & 39.44        & 42.38                 & 47.55                                   & N.A.                                   & 28.41        & 32.43         & 21.07                                   & 58.58                                   & { \textbf{66.96}}    \\
\multicolumn{1}{|c|}{\multirow{-5}{*}{\textbf{China}}}  & \multicolumn{1}{c|}{\textbf{SR}}   & 1.5489       & 1.5977                & 1.6938                                  & 0.6482                                 & 0.6922       & 1.0866        & 0.4658                                  & \textbf{2.1283}                         & { \textbf{2.3273}}   \\ \hline
\multicolumn{1}{|c|}{}                                  & \multicolumn{1}{c|}{\textbf{CR}}   & 7.30\%       & 8.33\%                & 13.81\%                                 & 8.65\%                                 & 6.91\%       & 9.85\%        & 9.44\%                                  & 14.35\%                                 & { \textbf{16.74\%}}  \\
\multicolumn{1}{|c|}{}                                  & \multicolumn{1}{c|}{\textbf{MER}}  & 10.18\%      & 12.74\%               & 18.96\%                                & 13.75\%                                & 7.28\%       & 10.77\%       & 11.28\%                                 & 18.24\% & \textbf{20.03\% }                                 \\
\multicolumn{1}{|c|}{}                                  & \multicolumn{1}{c|}{\textbf{MPB}}  & 10.03\%      & 11.30\%               & 9.98\%                                  & \textbf{6.79\%}                                 & 12.54\%      & 13.28\%       & 13.16\%                                 & 10.12\%                                 & 9.72\%   \\
\multicolumn{1}{|c|}{}                                  & \multicolumn{1}{c|}{\textbf{APPT}} & 16.99        & 18.86                 & 25.53   & N.A.                                   & 14.31        & 19.03         & 17.52                                   & 23.31                                   & \textbf{27.96}                                    \\
\multicolumn{1}{|c|}{\multirow{-5}{*}{\textbf{Indian}}} & \multicolumn{1}{c|}{\textbf{SR}}   & 0.4853       & 0.5506                & 0.8981                                  & 0.5709                                 & 0.4606       & 0.6470        & 0.6210                                  & 0.9323                                  & { \textbf{1.0839}}   \\ \hline
\multicolumn{1}{|c|}{}                                  & \multicolumn{1}{c|}{\textbf{CR}}   & 8.83\%       & 9.02\%                & 14.26\%                                 & 9.74\%                                 & 10.32\%      & 14.02\%       & { \textbf{18.60\%}} & 18.15\%                                 & 16.84\%                                  \\
\multicolumn{1}{|c|}{}                                  & \multicolumn{1}{c|}{\textbf{MER}}  & 8.83\%       & 9.02\%                & 17.68\%                                 & 14.68\%                                & 11.96\%      & 15.51\%       & 20.35\%                                 & 20.77\%                                 & { \textbf{22.59\%}}  \\
\multicolumn{1}{|c|}{}                                  & \multicolumn{1}{c|}{\textbf{MPB}}  & 18.59\%      & 17.18\%               & 18.24\%                                 & { \textbf{2.30\%}} & 30.65\%      & 39.78\%       & 38.75\%                                 & 39.22\%                                 & 27.54\%                                  \\
\multicolumn{1}{|c|}{}                                  & \multicolumn{1}{c|}{\textbf{APPT}} & 19.87        & 21.51                 & 32.41                                   & N.A.                                   & 21.38        & 26.12         & 35.77                                   & 33.61                                   & { \textbf{36.03}}    \\
\multicolumn{1}{|c|}{\multirow{-5}{*}{\textbf{UK}}}     & \multicolumn{1}{c|}{\textbf{SR}}   & 0.5572       & 0.5685                & 0.8789                                  & 0.6111                                 & 0.6455       & 0.8647        & { \textbf{1.1360}}  & 1.1094                                  & 1.0318                                   \\ \hline
\end{tabular}}
\end{table*}

Overall, our CLSTM-PPO model performs the best among all the benchmark models. Our model holds the first place among the four markets in both MER and APPT metrics, which fully demonstrates that the use of cascaded LSTM can effectively extract the potential time-series features in the market and thus enhance the profit mining ability of agent. For the cumulative return, it only underperforms in the UK market. This may be due to the limitations of the dataset: The agent does not have enough data to be fully trained and has a short trading time, which limits the performance of the agent trained by the deep reinforcement learning algorithm. From our experiment, when using deep reinforcement learning to build a stock trading strategy for day-frequency trading, more than 7 years of training data should be needed to allow agent to fully learn the different characteristics of the market. 

At the same time, our model performs average on the maximum pullback indicator. However, all benchmark models do not show an absolute advantage in this metric. Ensemble strategy is a relatively stable model in terms of pullback control. Therefore, the advantage of using multiple agents trained simultaneously is that it makes the model less likely to make big mistakes like Yang\cite{14}, but it also means less likely to obtain high returns, since risk and return are positively correlated in the investment world.

\section{Conclusion}
In this paper, we propose a PPO model using cascaded LSTM networks and compare the ensemble strategy in Yang\cite{14} as a baseline model in the US, Chinese, Indian and UK market, respectively. The results show that our model has a stronger profit-taking ability, and this feature is more prominent in the Chinese market. However, according to the risk-return criterion, our model is exposed to higher pullback risk while obtaining high returns. Finally, we believe that the strengths of our model can be more fully exploited in markets where the overall trend is smoother and less volatile, as returns can be more consistently accumulated in such an environment (such as the A-share market in China in recent years). This suggests that there is indeed a potential return pattern in the stock market, and the LSTM as a time-series feature extractor plays an active role. Also, it shows that the Chinese market is a suitable environment for developing quantitative trading. 

In the subsequent experiments, improvements can be made in the following aspects: (i) The amount of training data. The training of PPO requires a large amount of historical data to achieve good learning results, so expanding the amount of training data may help to improve the results. (ii) Reward function. Some improved reward functions for stock trading have emerged, which can enhance the stability of the algorithm. For example, we can refer to the Sharpe ratio approach to weigh risk and return to control pullback. The numerator is the mean of the return series and the denominator is the standard deviation of the return series. At the same time, if the reward signal has a very large or very small range, it can cause numerical instabilities during training. In this case, you can normalize the reward function to a smaller range (e.g., between -1 and 1) to make it more stable.





\end{document}